\documentstyle[aps,multicol,epsf,rotate]{revtex}
\newcommand{\ie}{{i.e.,\ }}

\newcommand{\la}{\alt}

\newcommand{\Ls}{L_{\!S}}
\newcommand{\Nf}{N_{\!f}}

\newcommand{\bfvf}{{\bf v}_{\!f}}
\begin{document}
%
%
\title{
{\sc XXI Int.\ Conf.\ on Low Temperature Physics in Prague, 
8-14 August 1996}\newline {\rm\small
Czechoslovak Journal of Physics {\bf 46}, 1005 (1996), Suppl. S2}\\[30pt]
Universality in Transport Processes of Unconventional Superconductors}
%
%
\author{M.\ J.\ Graf, S.-K.\ Yip, and J.\ A.\ Sauls}

%
\address{Dept.\ of Physics \& Astronomy, Northwestern University,
2145 Sheridan Rd., Evanston, IL 60208-3112, USA}

\maketitle
%
\begin{abstract}
 We show that some of the low temperature transport coefficients (e.g., 
electrical and thermal conductivities, viscosity and sound attenuation)
are {\it universal}, i.e., independent of the impurity concentration 
and phase shift for specific classes of unconventional superconductors.
  The existence of a universal limit depends on the symmetry of the 
order parameter and is achieved at low temperatures $k_B T \ll \gamma 
\ll \Delta_0$, where $\gamma$ is the bandwidth of the impurity induced 
Andreev bound states.  
  The density of states is finite at zero energy and leads to the 
re-appearance of the Wiedemann-Franz law deep in the superconducting phase 
for $k_B T \ll \gamma$.
  Our findings also show that impurity concentration studies at
low temperatures can distinguish between different order parameter 
symmetries.
\end{abstract}
\vspace{0.1in}{PACS: 74.25.Fy, 74.25.Ld, 74.72.-h, 74.70.Tx}

%
%
\begin{multicols}{2}
 
  We investigate the behavior of the heat current, the
electrical current, and the momentum current (transverse sound
attenuation) for an unconventional superconductor, \ie for an order 
parameter with reduced
symmetry for which gapless excitations exist even at zero temperature.
  Such superconducting states have been argued to both exist in the
cuprates and heavy fermion superconductors.  A leading candidate in
the cuprates for a tetragonal crystal structure (D$_{4h}$) is the
B$_{1g}$ state ($d_{x^2-y^2}$), a singlet state with lines of nodes
at the Fermi positions $p_{\!fx} = \pm p_{\!fy}$  \cite{Scalapino}.  
Similarly, the most
promising candidates in the heavy fermion metal UPt$_3$ are the 
two-dimensional orbital representations coupled to a symmetry breaking 
field.
  For UPt$_3$, which has a hexagonal crystal structure ($D_{6h}$), phase
diagram studies, and transport measurements lead to either an
even-parity, spin singlet $E_{1g}$, or an odd-parity, spin triplet
$E_{2u}$ pairing state. 
  In both cases the order parameter vanishes at
the Fermi surface on a line in the basal plane, $p_{\!fz}=0$, and at
points at the poles, $p_{\!fx}=p_{\!fy}=0$ \cite{Sauls}.

  Instead of examining the effects of the multi-sheeted Fermi surface 
on the heat current, charge current, and momentum current, we model the 
excitation spectrum by an excitation gap that opens at line and point 
nodes on the Fermi surface, and by the Fermi surface properties in the 
vicinity of the nodes (\ie the Fermi velocities, $\bfvf$, and the 
density of states, $\Nf$, near the nodes) \cite{Graf}.  
  Crystal symmetry determines the positions of the nodal regions of the
excitation gap on the Fermi surface, but not the prefactors for
the gap opening.  

  The low-temperature
behavior of the transport coefficients probes {\it lower-dimensional}
regions of the Fermi surface, $\epsilon\la \gamma\ll \Delta_0\ll E_f$,
where the excitation gap vanishes, and is less sensitive to the overall 
geometry of the Fermi surface.  
  The low-energy scale $\gamma$ is defined by the bandwidth of
the impurity induced Andreev bound states, and reflects
the formation of a novel metallic state deep in the superconducting
phase;  for strong scattering $\gamma \propto \sqrt{\Gamma_0\Delta_0}$,
where $\Gamma_0$ is the (elastic) normal-state scattering rate 
$\hbar/2\tau(0)$.
  We parametrize the nodal regions of the gap with a minimal set of nodal 
parameters, and attempt to fit these parameters in order to achieve 
accurate low temperature limits for the different transport
coefficients along the principal axes of the crystal. 
  Thus, our order parameter model depends on angle
(${\bf p}_{\!f}$) and the nodal parameters (\{$\mu_i$\}), 
$\Delta({\bf p}_{\!f};\{\mu_i\})$.
  The advantage of this approach is that we can quantitatively determine
the phase space contributing to the low temperature transport 
coefficients and then examine in more detail the effects of impurity 
scattering and order parameter symmetry on the current response
\cite{Graf}, without having to know the overall shape of the Fermi 
surface or basis functions.

  The number of nodal parameters is fixed by
the minimal number of symmetry unrelated point and
line nodes.  These parameters define the slope or curvature
of the gap near a line or point node in a spherical coordinate 
system  (uniaxial anisotropy is included by mapping an ellipsoidal 
Fermi surface onto a sphere).
  Specifically, for the E-rep models of UPt$_3$ we parametrize
the gap in the vicinity of the equatorial line node by
$|\Delta(\theta)|\approx
\mu\Delta_0|\frac{\pi}{2}-\theta|$,
while near the poles $|\Delta(\theta)|\approx 
\mu_n\Delta_0|\theta|^{n}$,
where the internal phase winding number is $n=1$ for $E_{1g}$ and 
$n=2$ for $E_{2u}$.
  This parametrization allows us to adjust {independently} 
the opening of the gap at the line and point nodes in order to
describe the low-energy excitation spectrum.
  Note that the crucial difference between the $E_{1g}$ and $E_{2u}$
state lies in the opening of the gap 
at the polar point nodes.  A similar parametrization yields for the 
$d_{x^2-y^2}$-wave order parameter model in the cuprates with
a cylindrical Fermi surface: $\Delta(\varphi)\approx 
\mu\Delta_0(\frac{\pi}{4}-\varphi)$, 
$\varphi\approx \pi/4$.

  We calculated in linear response (including vertex corrections), 
and in the long wavelength
limit ($q \to 0$), for sufficiently low temperatures and external
frequencies ($k_B T, \hbar\omega \ll \gamma \ll \Delta_0$) the
transport coefficients in the limit of weak and strong impurity
scattering.  The asymptotic values, which we derived for the electrical
and thermal conductivity, as well as for the transverse viscosity 
$\boldmath{\eta}$ and ultrasound attenuation $\boldmath{\alpha}$,
are listed in Table~I.
  Here, we restricted ourselves to the hydrodynamic limit and
orientations of the wavevector $\bf q$ and polarization 
$\boldmath{\varepsilon}$, such that $\alpha_{i j} = 
({q^2}/{\varrho\,c_s})\eta_{i j, i j}\hat\varepsilon_i\,\hat q_j$,
where $\varrho$ is the mass density and $c_s$ the speed of sound.

%
%

\vspace{10pt}
\noindent
{\small Table I.\
Asymptotic low-$T$ and low-$\omega$ values of the transport 
coefficients 
for three different pairing states, where we have used
$\sigma_{00}=e^2\, N_{\!f} v^2_{\!f}\tau_{00}$,
$\kappa_{00}=(\pi^2/3)k_B^2 T\, N_{\!f} v^2_{\!f}\tau_{00}$, and
$\alpha_{00}= (q^2/4\varrho\,c_s)p^2_{\!f}\, N_{\!f} v^2_{\!f}\tau_{00}$
with an effective $\tau_{00}=\hbar/2\mu\Delta_0(0)$.  }
\vspace{0pt}
\begin{center}
\begin{tabular}{cccc}
\noalign{\smallskip}\hline\hline
	transport coeff.
  &	$d_{x^2-y^2}$
  &	$E_{1g}$
  &	$E_{2u}$
\\ \hline
    $\sigma_{xx}(T\!\to\! 0)/\sigma_{00}$
  & $\displaystyle {4}/{\pi}$
  & $\displaystyle {1}$
  & $\displaystyle {1}$
\\[1.0ex]
    $\sigma_{zz}(T\!\to\! 0)/\sigma_{00}$
  & ---
  & $\displaystyle {2\mu\gamma}/({\mu_1^2\Delta_0})$
  & $\displaystyle {\mu}/{\mu_2}$
\\[1.0ex]
    $\kappa_{xx}(T\!\to\! 0)/\kappa_{00}$
  & $\displaystyle {4}/{\pi}$
  & $\displaystyle {1}$
  & $\displaystyle {1}$
\\[1.0ex]
    $\kappa_{zz}(T\!\to\! 0)/\kappa_{00}$
  & ---
  & $\displaystyle {2\mu\gamma}/({\mu_1^2\Delta_0})$
  & $\!\displaystyle {\mu}/{\mu_2}\!$
\\[1.0ex]
    $\alpha_{xy}(T\!\to\! 0)/\alpha_{00}$
  & $\!\displaystyle {2}/{\pi}\!$
  & $\!\displaystyle {1}\!$
  & $\displaystyle {1}$
\\[1.0ex]
    $\alpha_{xz}(T\!\to\! 0)/\alpha_{00}$
  & ---
  & $\displaystyle \frac{8\mu}{1+2\mu^2}\frac{\Gamma_0}{\Delta_0}^\star$
  & $\displaystyle \frac{2\mu}{\mu_2^2}\frac{\gamma}{\Delta_0}$
\\
\noalign{\smallskip}\hline\hline
\end{tabular}
\end{center}
\vspace{-1pt}
{\small $^\star$ In the strong scattering limit including vertex
corrections.}
\vspace{10pt}

  The transport coefficients in the basal plane (or CuO$_2$ planes) are 
{\it universal}, \ie independent of the impurity concentration and
scattering phase shift. Furthermore, the in-plane results do 
not distinguish between the two E-rep models in UPt$_3$.  
  Estimates of these coefficients are in good
agreement with experiments \cite{Graf}.
  However, transport measurements along the crystal 
$c$ axis {\it can} distinguish between an $E_{1g}$ or $E_{2u}$ 
pairing state.
  For the electrical and thermal conductivity the $E_{2u}$ 
state leads to a universal value at low temperatures, independent 
of impurity scattering, while the $E_{1g}$ state has a non-universal 
value, strongly dependent on impurity scattering.

  Independent of the universality of the transport coefficients 
we find that the Wiedemann-Franz law is re-established at low 
temperatures 
$k_B T \la \gamma$, due to the finite density of states (DOS) at zero energy.
  At finite temperatures the Lorenz ratio $L(T)$ deviates significantly
from Sommerfeld's value $\Ls = \frac{\pi^2}{3} (k_B/e)^2$ even for 
elastic scattering,  because of the different coherence factors in the 
electrical and thermal conductivity, and the energy dependence of the DOS.  
  The Lorenz ratio depends sensitively on the scattering phase shift.
  For strong scattering the ratio $L(T)/\Ls$ is typically larger than 
one, whereas for weak scattering (Born) it is less than one \cite{Graf}.
  This is a very robust feature, and remains true when we include 
inelastic scattering, modeled by a phenomenological temperature dependent 
relaxation time, as shown in Fig.~1.

\vspace{5pt}
\noindent
\begin{minipage}{80mm}
\centerline{\epsfysize=75mm \epsfxsize=47mm
\rotate[r]{\epsfbox{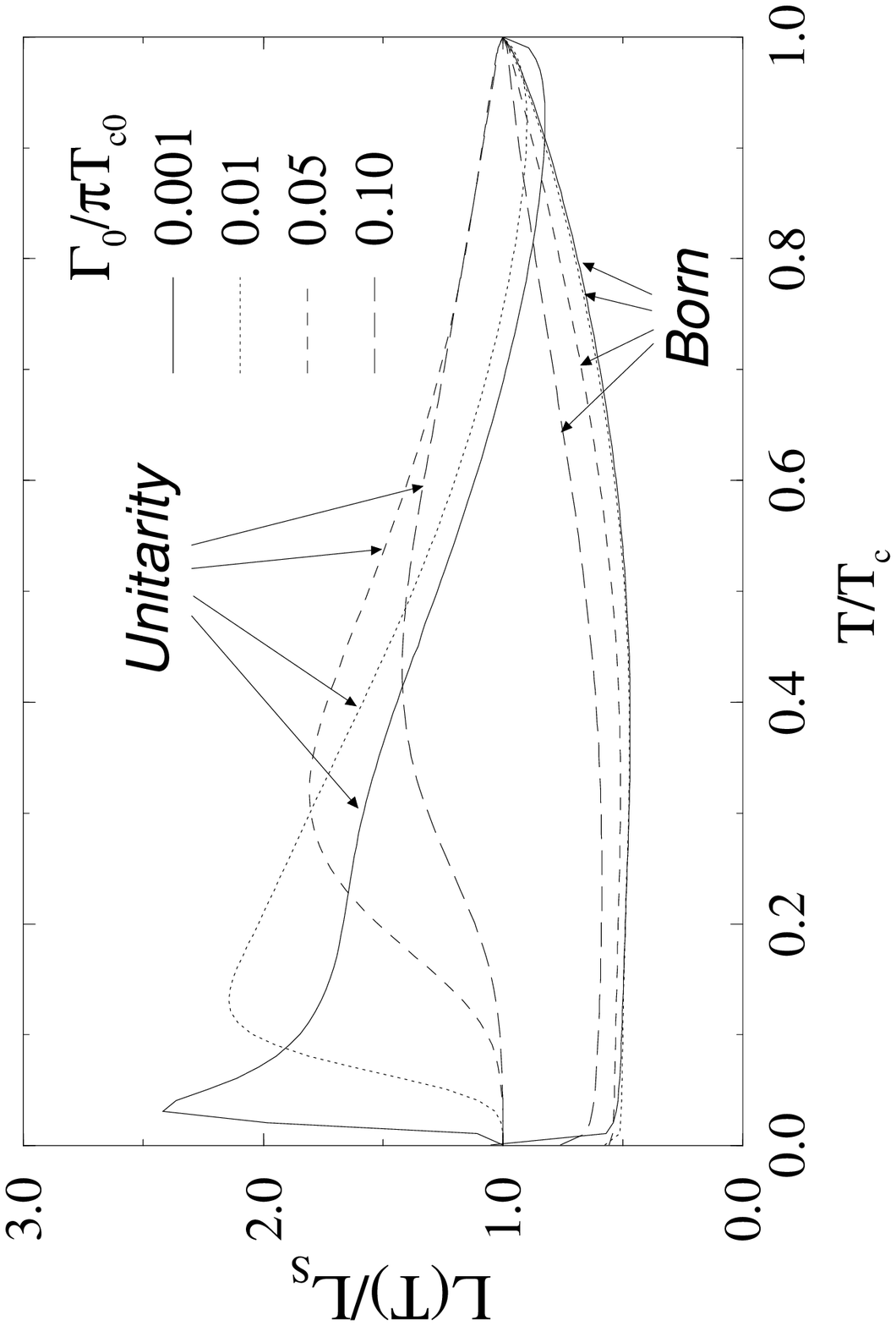}}}
\vspace{-5pt}
\noindent
{\small Fig.~1 \
Lorenz ratio $L(T)$ for an $E_{2u}$ state for weak (Born) and strong 
(unitarity) scattering and various phenomenological scattering rates 
$\Gamma(T) = \Gamma_0 (1+T^2/T^2_c)$.}
\end{minipage}
\vspace{7pt}

  As a result, measurements of transport coefficients
along different crystal axes at very low temperatures and for various 
impurity concentrations will distinguish between the $E_{1g}$ and
$E_{2u}$ pairing models in UPt$_3$, and also elucidate the pairing 
state in the cuprates.

\acknowledgments
This research was supported by the NSF through
the Science and Technology Center for Superconductivity (DMR 91-20000),
and the Northwestern University Materials Science Center 
(DMR 91-20521).

%
%

\end{multicols}
\end{document}